\documentclass[preprint,prd,nofootinbib,tightenlines,amsmath,amssymb]{revtex4}
\usepackage{enumerate}
\usepackage{multirow}
\usepackage[utf8]{inputenc}
\usepackage{anysize}
\usepackage{textcomp}
\usepackage{epsfig}
\usepackage{graphics,color} 
\usepackage{verbatim}
\usepackage{afterpage}
\usepackage{amsmath}    % need for subequations
\usepackage{soul}
\usepackage{xcolor,cancel}
\DeclareUnicodeCharacter{00A0}{ }

\usepackage{blindtext}

\usepackage{mathtools}

\catcode`\@=11
\def\sla@#1#2#3#4#5{{%
 \setbox\z@\hbox{$\m@th#4#5$}%
 \setbox\tw@\hbox{$\m@th#4#1$}%
 \dimen4\wd\ifdim\wd\z@<\wd\tw@\tw@\else\z@\fi
 \dimen@\ht\tw@
 \advance\dimen@-\dp\tw@ \advance\dimen@-\ht\z@
 \advance\dimen@\dp\z@
 \divide\dimen@\tw@ \advance\dimen@-#3\ht\tw@
 \advance\dimen@-#3\dp\tw@ \dimen@ii#2\wd\z@
 \raise-\dimen@\hbox to\dimen4{%
 \hss\kern\dimen@ii\box\tw@\kern-\dimen@ii\hss}%
 \llap{\hbox to\dimen4{\hss\box\z@\hss}}}}

\def\cpto{\mathrel {\vcenter {\baselineskip 0pt \kern 0pt
    \hbox{$H_{r.f.}$} \kern 0pt \hbox{$\longrightarrow$} }}}
\def\slashed#1{%
 \expandafter\ifx\csname sla@\string#1\endcsname\relax
{\mathpalette{\sla@/00}{#1}}
\fi}
\def\declareslashed#1#2#3#4#5{%
 \expandafter\def\csname sla@\string#5\endcsname{%
#1{\mathpalette{\sla@{#2}{#3}{#4}}{#5}}}}
 \catcode`\@=12
\declareslashed{}{/}{.08}{0}{D}
 \declareslashed{}{/}{.1}{0}{A}
 \declareslashed{}{/}{0}{-.05}{k}
 \declareslashed{}{/}{.1}{0}{\partial}
 \declareslashed{}{\not}{-.6}{0}{f}

\def\lsim{\mathrel {\vcenter {\baselineskip 0pt \kern 0pt
    \hbox{$<$} \kern 0pt \hbox{$\sim$} }}}
\def\gsim{\mathrel {\vcenter {\baselineskip 0pt \kern 0pt
    \hbox{$>$} \kern 0pt \hbox{$\sim$} }}}

\newcommand{\bea}{\begin{eqnarray}}
\newcommand{\eea}{\end{eqnarray}}

\def\ln#1{U^{L*}_{#1 N}}

\def\lns#1{U^{L}_{#1 N}}

\begin{document}

\baselineskip=15pt
\preprint{}

\title{Are the B-anomalies evidence for heavy neutrinos?}

\author{Xiao-Gang He$^{1,2,3}$\footnote{Electronic address: hexg@phys.ntu.edu.tw} and German Valencia$^{4}$\footnote{Electronic address: German.Valencia@monash.edu }}
\affiliation{
$^{1}$Physics Division, National Center for Theoretical Sciences, Hsinchu, Taiwan 30013.\\
$^{2}$T-D Lee Institute,Department of Physics and Astronomy, Shanghai Jiao Tong University, Shanghai.\\
$^{3}$Department of Physics, National Taiwan University, Taipei. \\
$^{4}$School of Physics and Astronomy, Monash University, 3800 Melbourne Australia.}

\date{\today}

\vskip 1cm
\begin{abstract}

The existing anomalies appearing in decays of the form $b\to s \ell^+ \ell^-$ constitute a possible hint for new physics. We point out that modifications to the SM results due to heavy neutrinos could account for the observed deviations while satisfying existing constraints from lepton flavor violating processes. The required mixing angle, however, is an order of magnitude larger than suggested by recent global fits to lepton flavor conserving processes. We frame our discussion in terms of a Type-I seesaw model, but it can be made more general.

\end{abstract}

\pacs{PACS numbers: }

\maketitle

\section{Introduction}

Experimental data have suggested anomalies in the flavour-changing neutral current (FCNC) process $b\to s \mu^+ \mu^-$ for some time now \cite{Aaij:2013qta,Aaij:2014pli,Aaij:2015oid,Aaij:2013aln,Aaij:2015esa}. At the same time it appears that the related mode with electrons instead of muons $b\to s e^+ e^-$, is consistent with the standard model (SM) expectations \cite{Aaij:2015dea}. A particularly interesting discrepancy between experiment and the SM occurs in the ratios $R_K,R_K^\star = B(B\to K(K^\star)\mu^+\mu^-)/B(B\to K (K^\star)e^+e^-)$ \cite{Aaij:2014ora,Abdesselam:2016llu,Wehle:2016yoi}, where lepton universality appears to be violated.

As expected, the anomalies in the $b\to s \ell^+ \ell^-$ measurements have received considerable attention in the literature and multiple models have been put forward as possible new physics explanations~\cite{models}. There are also several model independent analyses of these experimental results in the form of global fits to the Wilson coefficients of the relevant low energy effective Hamiltonian ~\cite{Capdevila:2017bsm}. 

One of the scenarios preferred by these global fits affects primarily the $C_9$ and $C_{10}$ Wilson coefficients. These coefficients are defined by the operators,
\begin{eqnarray}
&&
{\cal H}_{\rm eff}
= -\frac{4G_F}{ \sqrt{2}}V_{tb}V^*_{ts}
\left( C_9 {\cal O}_9 + C_{10}{\cal O}_{10} \right)\;,\nonumber\\
&&
{\cal O}_9 = \frac{e^2}{16\pi^2} \left( \bar s \gamma_\mu P_Lb \right)
\left( \bar \ell \gamma^\mu \ell \right) \;,\;\;
{\cal O}_{10} = \frac{e^2}{16\pi^2} \left( \bar s \gamma_\mu P_Lb \right)
\left( \bar \ell \gamma^\mu\gamma_5 \ell \right) \;,
\end{eqnarray}
where $P_L = (1-\gamma_5)/2$ and, in the absence of flavour universality, $C_{9\mu,10\mu}\neq C_{9e,10e}$. The SM predicts that $C_{9,10}$ are approximately the same for all leptons with $C_9^{\rm SM} \approx 4.1$, and $C_{10}^{\rm SM} \approx -4.1$. The model discussed in this paper will introduce corrections to these coefficients with the pattern $C^{NP}_9(M_W)=-C^{NP}_{10}(M_W)$ and therefore our benchmark will be the best fit with $C_{9\mu}^{NP} =-C_{10\mu}^{NP}$ in the 1$\sigma$ range $[-0.73,-0.48]$ found by Ref.~\cite{Capdevila:2017bsm}. 

\section{The model}

In Type I Seesaw models~\cite{seesaw} there are three light and N heavy
neutrinos and the general mass term for the neutrinos can be written as
\begin{eqnarray}
{\cal L}_M = -\bar L_L Y_\nu \phi \nu_R -\frac{1}{ 2} \bar \nu^{c}_R M_R \nu^{}_R
 + h.c.
\end{eqnarray}
where $\phi$ is the usual Higgs doublet with a vacuum expectation value $\langle \phi\rangle = v/\sqrt{2}$.
The neutrino mass matrix then takes the following form in the $(\nu^c_L, \nu_R)$ basis
\begin{eqnarray}
M_\nu = \left ( \begin{array}{rr}
0&\frac{v}{ \sqrt{2}}Y_\nu\\
 \frac{v}{\sqrt{2}}Y^T_\nu& M_R
\end{array}
\right ),
\end{eqnarray}
which is a symmetric matrix that can be diagonalized by the transformation 
\begin{eqnarray}
\tilde U^T M^\nu \tilde U = \hat M^\nu.
\end{eqnarray}
$\tilde U$ is a unitary matrix and the diagonal neutrino mass matrix is then $\hat M^\nu = diag(m_1,
m_2,m_3, M_4, \cdots, M_{3+N})$ with $m_i$ and $M_i$  the light and
heavy mass eigenvalues respectively. We will denote the heavy neutral mass eigenstates by $N$.
$\tilde U$ is a $(3+N)\times (3+N)$ matrix which can be written as two block $3\times (3+N)$ matrices
\begin{eqnarray}
\tilde U = \left ( \begin{array}{l} \tilde U^{L}\\ \tilde U^R
\end{array}
\right ),
\end{eqnarray}
In order to accommodate the known neutrino oscillation data, which shows that there are at least two massive light neutrinos, $N$ should be equal to or larger than two.

The charged current interaction between the $W$-boson and quarks and leptons in the weak interaction basis is given by
\begin{eqnarray}
{\cal L}_{CC} &=& -\frac{g}{ \sqrt{2}}W^\mu \bar L \gamma_\mu \nu -\frac{g}{ \sqrt{2}}W^\mu \bar U \gamma_\mu P_L D + h.c.
\end{eqnarray}
where $L = (e,\;\;\mu,\;\;\tau)^T$, $\nu =
(\nu_e,\;\;\nu_\mu,\;\;\nu_\tau)^T$, $U = (u,\;\;c,\;\;t)^T$,
and $D = (d,\;\;s,\;\;b)^T$. If there are no right-handed W-boson interactions, the heavy right-handed neutrinos are not connected to the charged leptons by the $W$ boson. However, the left-handed neutrinos will have heavy neutrino components  
in the mass eigenstate basis and the charged current becomes 
\begin{eqnarray}
{\cal L}_{CC} &=& -\frac{g}{ \sqrt{2}}W^\mu \bar \ell^m \gamma_\mu P_L U^{L*}\nu^m
-\frac{g }{\sqrt{2}}W^\mu \bar U^m_i \gamma_\mu P_L V D^m + h.c.\nonumber\\
&=&-\frac{g}{ \sqrt{2}}W^\mu\sum_{\ell=1}^3\bar \ell^m_\ell\gamma_\mu P_L U^{L*}_{\ell j} \nu^m_j
-\frac{g }{ \sqrt{2}}W^\mu \bar U^m_i \gamma_\mu P_L V_{ij} D^m_j + h.c.
\end{eqnarray}
Here we have introduced the matrix  $U^{L*}_{\ell j} = \sum_{i=1}^3
S^{\dagger L}_{\ell i}\tilde U^*_{ij}$  with  $\ell = e,\mu,\tau$. $S^{L}$ is the matrix that diagonalizes the
left-handed charged lepton mass matrix: $\ell^m_L = S^L
\ell_L$ and $V = V_{KM}$ is the standard Kobayashi-Maskawa (KM) matrix.

In what follows we will  drop the superscript ``m''
from the fermion fields and always refer to mass eigenstates. We will write the $3\times (3+N) $,  $U^{L} = (S^{L})^T\tilde U^L$, matrix  in the following form
\begin{eqnarray}
U^L & =& \left ( \begin{array}{llllll}
U^L_{e 1}&U^L_{e 2}&U^L_{e 3}&U^L_{e 4}&...&U^L_{e 3+N}\\
U^L_{\mu 1}&U^L_{\mu 2}&U^L_{\mu 3}&U^L_{\mu 4}&...&U^L_{\mu 3+N}\\
U^L_{\tau 1}&U^L_{\tau 2}&U^L_{\tau 3}&U^L_{\tau 4}&...&U^L_{\tau 3+N}\end{array}
\right ),
\end{eqnarray}
and note that it satisfies the unitarity condition 
\begin{eqnarray}
\sum_{j=1}^{3+N} U^{L*}_{\ell j} U^L_{\ell^\prime j}=\delta_{\ell\ell^{\prime}}.
\end{eqnarray}

\section{Low energy effective Lagrangian}

\begin{figure}[!htb]
\begin{center}
\includegraphics[width=6cm]{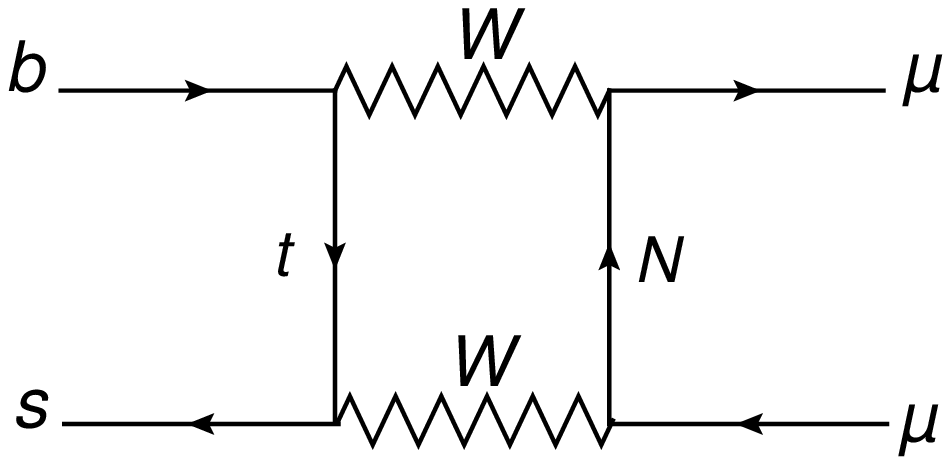}
\end{center}
\caption{Box diagram responsible for the process $b\to d_j \bar \ell \ell^\prime$. }
\label{f:box}
\end{figure}

The model is particularly simple, as the only new contribution to $B$ decay arises from the box diagram depicted in Figure~\ref{f:box} (plus associated diagrams involving would-be Golstone bosons).

\begin{figure}[!htb]
\begin{center}
\includegraphics[width=8cm]{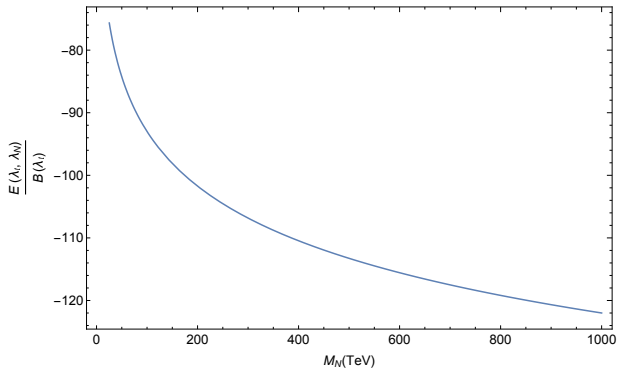}
\end{center}
\caption{Inami-Lim function $E(\lambda_t,\lambda_N)$ for physical top-quark mass as a function of heavy neutrino mass in TeV, normalized to the SM box function $B(\lambda_t)$. }
\label{f:inami}
\end{figure}

These diagrams have been calculated before for the case of lepton flavor violating (LFV) B decays and the result is known   \cite{Langacker:1988vz,Gagyi-Palffy:1994hx,He:2004wr,Fujihara:2005uq}. In our case we must be careful not to discard the terms that vanish due to the GIM mechanism on the neutrino side for  LFV processes, but do not vanish for lepton flavor conserving processes. We find the contribution to the effective Lagrangian for $b\to d_j \bar \ell \ell^\prime$ at the $M_W$ scale to be,
\begin{equation}
{\cal L}= -\frac{G_F}{\sqrt{2}} \frac{\alpha }{ 2 \pi
s_W^2}\sum_{i=u,c,t}\sum_{\alpha=1 \cdots N+3}
V^{*}_{id_j}V_{ib} U^{L\star}_{\ell \alpha}U^L_{\ell^\prime \alpha}
\left(4B(\lambda_i)+E(\lambda_i,\lambda_a)\right) \bar\ell \gamma_\mu
P_L \ell^\prime
\bar d_j \gamma^\mu P_L  b
\label{bleftnu}
\end{equation}
where $d_j$ refers to a $d$ or an $s$ quark,  $\lambda_i=m_i^2/m_W^2$, and the Inami-Lim functions  \cite{Inami:1980fz} $B(\lambda_i)$ and $E(\lambda_i,\lambda_\alpha)$ are given by,\footnote{Note that $E(\lambda_i,\lambda_\alpha)=-E_L(\lambda_t,\lambda_N)/4$ in the notation of Ref.~\cite{He:2004wr}. } 
\begin{eqnarray}
B(\lambda_i)&=& \frac{1}{4}\left(\frac{\lambda_i}{1-\lambda_i}+\frac{\lambda_i\log(\lambda_i)}{(1-\lambda_i)^2}\right) \nonumber \\
E(\lambda_i,\lambda_a)&=&\lambda_i\lambda_a\left\{ -\frac{3}{4}\frac{1}{(1-\lambda_i)(1-\lambda_a)}+\left[\frac{1}{4}-\frac{3}{2}\frac{1}{\lambda_i-1}-\frac{3}{4}\frac{1}{(\lambda_i-1)^2}\right]\frac{\log\lambda_i}{\lambda_i-\lambda_a}\right.\nonumber \\
&+&\left. \left[\frac{1}{4}-\frac{3}{2}\frac{1}{\lambda_a-1}-\frac{3}{4}\frac{1}{(\lambda_a-1)^2}\right]\frac{\log\lambda_a}{\lambda_a-\lambda_i}\right\}.
\end{eqnarray}
$B(\lambda_i)$ is just the usual function that reproduces the SM box diagram contribution to $b\to d_j \bar \ell \ell$ \cite{Buchalla:1995vs}. The  new term is given by $E(\lambda_i,\lambda_a)$ and it subtracts from the SM as illustrated in Figure~\ref{f:inami}.

Neglecting for simplicity the contribution from the charm-quark intermediate state, our result in Eq.~\ref{bleftnu} implies that
\begin{eqnarray}
C^{NP}_{9}(M_W)=-C^{NP}_{10}(M_W)= -\frac{1}{4s_W^2}\sum_N U^{L\star}_{\mu N}U^L_{\mu N}E(\lambda_t,\lambda_N).
\end{eqnarray}

As shown in Ref.~\cite{He:2004wr} the main constraints on the mixing angles $U^L_{\ell N}$ arise from radiative lepton decay and the corresponding low energy effective Lagrangian for this type of process is given by
\begin{eqnarray}
{\cal L} &=& 4 \frac{G_F }{ \sqrt{2}} \frac{e }{ 16\pi^2} F^{\mu\nu}
\sum_N 
 \ln{\ell} \lns{\ell^\prime} F(\lambda_N)\
m_{\ell^\prime} \bar\ell \sigma_{\mu\nu} P_R \ell^\prime 
\label{lplgops}
\end{eqnarray}
where $F^{\mu\nu}$ is the electromagnetic field strength tensor and
the Inami-Lim function in this case is
\begin{eqnarray}
F(\lambda_N) &=& \left[ \frac{3 \lambda_N^3\log\lambda_N}{
4(1-\lambda_N)^4} + \frac{2\lambda_N^3+5\lambda_N^2-\lambda_N}{ 8(1-\lambda_N)^3}\right],
\label{lplgcoef}
\end{eqnarray}

\section{Results and conclusion}

Using Eq.~\ref{lplgops} and Eq.~\ref{bleftnu} we can update the constraints on $U^L_{\ell N}$ arising from radiative lepton decay and LFV $B$ decay with the experimental limits given in Table~\ref{t:bres2m}.
\begin{table}[htp]
\caption{Summary of current experimental bounds for $\ell\to\ell^\prime \gamma$ and $B\to \ell^{\pm}\ell^{\prime \mp}$ taken from the Particle Data Book \cite{Olive:2016xmw} .} 
\begin{center}
\begin{tabular}{|c|c||c|c|}\hline
Process& limit  &Process& limit   \\ 
\hline 
$B(\mu \to e \gamma)$& $  4.2 \times 10^{-13}$ & $B\rightarrow e^\pm\mu^\mp$& $ 2.8\times 10^{-9}$ \\
$B(\tau \rightarrow e \gamma)$& $  3.3 \times 10 ^{-8}$ &$B\rightarrow e^\pm\tau^\mp$&  $ 2.8\times 10^{-5}$\\
$B(\tau \rightarrow \mu \gamma)$& $   4.4 \times 10 ^{-8}$ &$B\rightarrow \mu^\pm\tau^\mp$&  $  2.2\times 10^{-5}$\\ 
$K_L\rightarrow e^\pm\mu^\mp$& $ 4.7\times 10^{-12}$&$B_s\rightarrow e^\pm\mu^\mp$& $ 1.1\times 10^{-8}$ \\  
\hline
\end{tabular}
\end{center}
\label{t:bres2m}
\end{table}

\begin{figure}[!htb]
\begin{center}
\includegraphics[width=6cm]{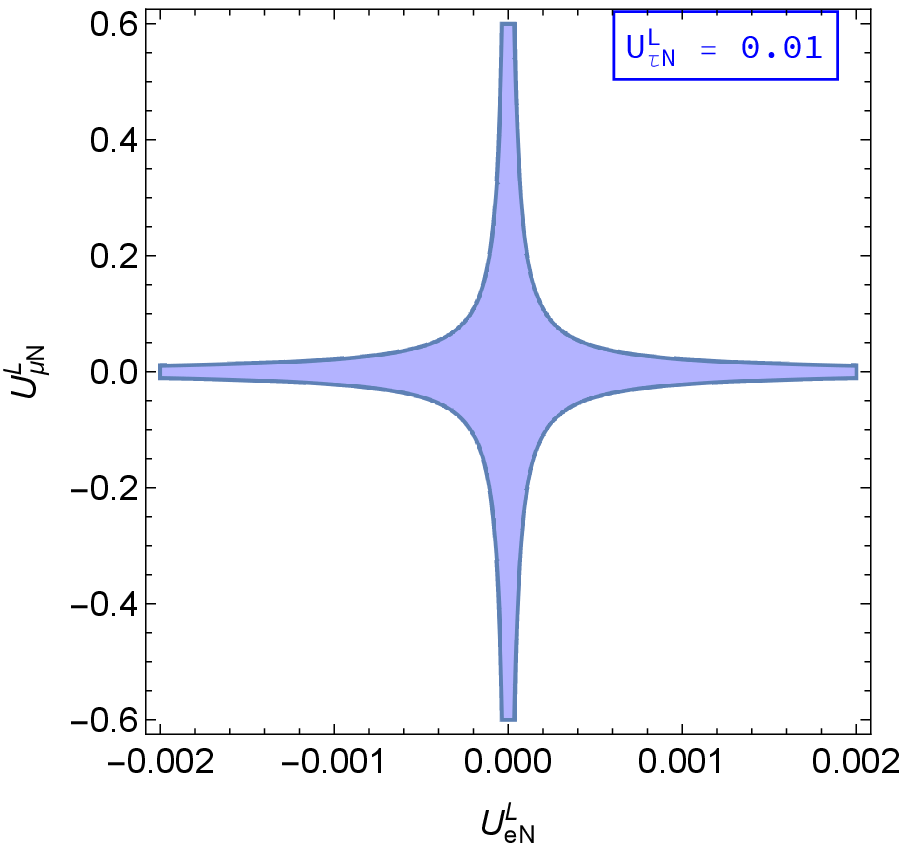}\hspace{1cm}\includegraphics[width=6cm]{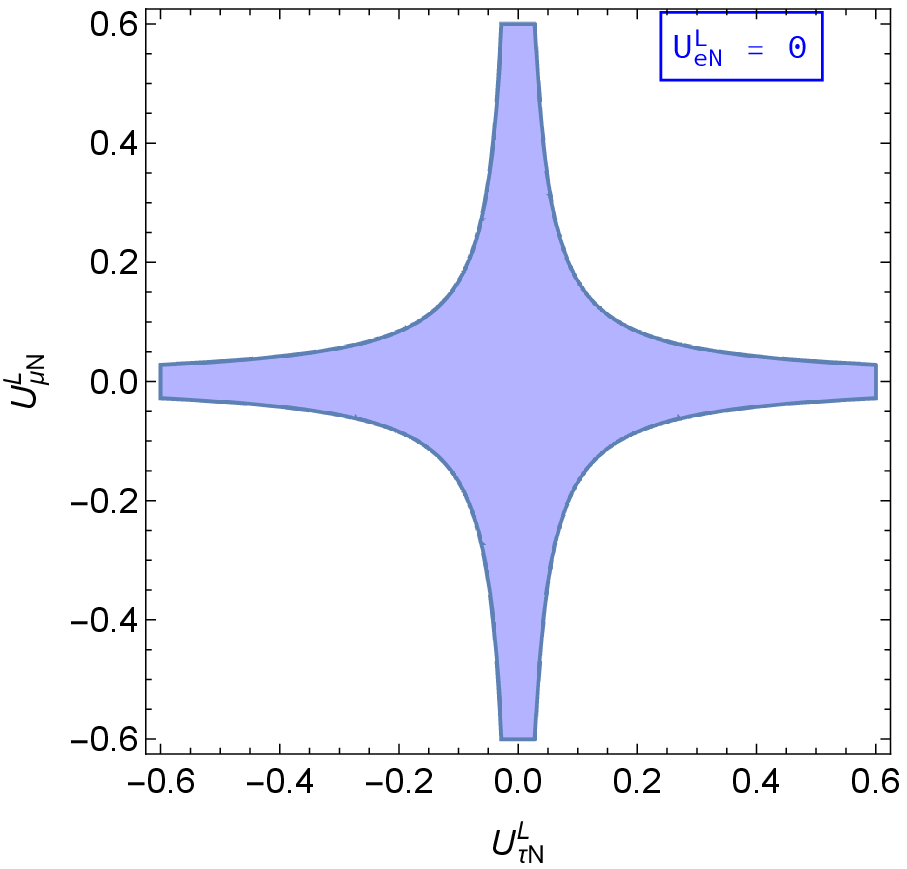}
\end{center}
\caption{Constraints on $U^L_{\mu N}$ for $M_N =2$~TeV in two illustrative scenarios. }
\label{f:cons}
\end{figure}

For simplicity we now consider  the contribution of only one heavy neutrino and we find that its mass is not significantly constrained. Taking, for example, $M_N = 2$~TeV and assuming that all the elements $U^L_{\ell N}$ are real, we illustrate two scenarios in Figure~\ref{f:cons}. 
Although not necessary, it is possible to choose $U^L_{eN}=0$, which will automatically remove any  constraints from LFV processes involving electrons, and we show this case in the right panel. The constraints in this case arise exclusively from $\tau \rightarrow \mu \gamma$. 

\begin{figure}[!htb]
\begin{center}
\includegraphics[width=8cm]{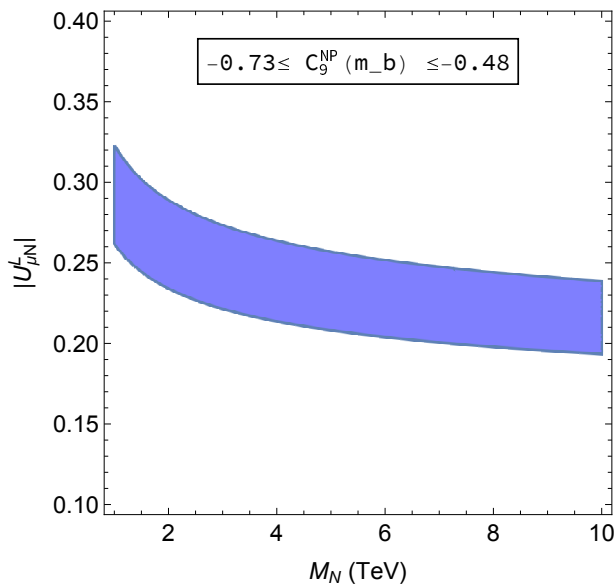}
\end{center}
\caption{Region of $U^L_{\mu N}-M_N$~(TeV) parameter space that satisfies the LFC constraints and produce a $C_{9\mu}^{NP}\approx -C_{10\mu}^{NP}$ in the range preferred by the global fit of Ref.~\cite{Capdevila:2017bsm}. }
\label{allowed}
\end{figure}

The parameter space that both, reproduces the best fit scenario of Ref.~\cite{Capdevila:2017bsm} at 1$\sigma$,  $C_{9\mu}^{NP}(m_b) \approx -C_{10\mu}^{NP}(m_b) \sim [-0.73,-0.48]$, and satisfies the LFV constraints is shown in  Figure~\ref{allowed}.\footnote{ Notice that this is only approximate for  $C_{9\mu}^{NP}(M_W) = -C_{10\mu}^{NP}(M_W)$ as the QCD running changes $C_9$ but not $C_{10}$ \cite{Buchalla:1995vs}. }
To accommodate the B-anomalies  with heavy neutrino masses in the TeV range, would thus require $U^L_{\mu N}\sim 0.3$. This would produce a contribution to the muon $g-2$ of $a_\mu = -5.7\times 10^{-10}$ which is at the level of the error in the measurement and below the current anomaly. At the same time, the large mixing angle needed is at odds with recent global fits \cite{Antusch:2006vwa,Fernandez-Martinez:2016lgt}.

Experimental anomalies in the tree-level dominated semileptonic decay of $b$ to $\tau$-leptons have also been reported \cite{Bozek:2010xy,Lees:2012xj,Aaij:2015yra}. These ones, 
$R(D)=B(B\to D\tau^- \bar\nu_\tau)/B(B\to D\ell^- \bar\nu_\ell)$ and $R(D^\star)$ (where a $D^\star$ replaces the $D$) cannot be explained by the mechanism described in this paper. Interestingly, however, there exist possible explanations involving additional light neutrinos for this case \cite{He:2012zp,Cvetic:2017gkt}.

\begin{acknowledgments}

X-G He was supported in part by MOE Academic Excellent Program (Grant No.~102R891505) and MOST of ROC (Grant No.~MOST104-2112-M-002-015-MY3), and in part by NSFC (Grant Nos.~11175115 and 11575111) and Shanghai Science and Technology Commission (Grant No.~11DZ2260700) of PRC.

\end{acknowledgments}


\begin{thebibliography}{999}

%%%%%%% experiment

\bibitem{Aaij:2013qta}
  {LHCb} Collaboration,
 % ``Measurement of Form-Factor-Independent Observables in the Decay $B^{0} \to K^{*0} \mu^+ \mu^-$,''
  PRL\  {\bf 111} (2013) 191801,
  arXiv:1308.1707 [hep-ex].
  %%CITATION = ARXIV:1308.1707;%%
  
\bibitem{Aaij:2014pli}
  {LHCb} Collaboration,
 % ``Differential branching fractions and isospin asymmetries of $B \to K^{(*)} \mu^+ \mu^-$ decays,''
  JHEP {\bf 1406} (2014) 133,
  arXiv:1403.8044 [hep-ex].
  %%CITATION = ARXIV:1403.8044;%%  
  
 %\cite{Aaij:2015oid}
\bibitem{Aaij:2015oid} 
  {LHCb} Collaboration,
  %``Angular analysis of the $B^{0}\rightarrow K^{*0}\mu^{+}\mu^{-}$ decay,''
  arXiv:1512.04442 [hep-ex].
  %%CITATION = ARXIV:1512.04442;%%
  %1 citations counted in INSPIRE as of 18 Jan 2016 
  
\bibitem{Aaij:2013aln}
  {LHCb} Collaboration,
%  ``Differential branching fraction and angular analysis of the decay $B_s^0\to\phi\mu^{+}\mu^{-}$,''
  JHEP {\bf 1307} (2013) 084,
  arXiv:1305.2168 [hep-ex].
  %%CITATION = ARXIV:1305.2168;%%  
  
  \bibitem{Aaij:2015esa}
  {LHCb} Collaboration,
%{``Angular analysis and differential branching fraction of the decay $B^0_s\to\phi\mu^+\mu^-$,''}
  arXiv:1506.08777 [hep-ex].
  %%CITATION = ARXIV:1506.08777;%%
  
\bibitem{Aaij:2015dea}
  {LHCb} Collaboration,
%  ``Angular analysis of the B$^{0}$ ? K$^{*0}$ e$^{+}$ e$^{?}$ decay in the low-q$^{2}$ region,''
  JHEP {\bf 1504} (2015) 064,
  arXiv:1501.03038 [hep-ex].
  %%CITATION = ARXIV:1501.03038;%%
  
 \bibitem{Aaij:2014ora}
  {LHCb} Collaboration,
 % ``Test of lepton universality using $B^{+}\rightarrow K^{+}\ell^{+}\ell^{-}$ decays,''
  Phys.\ Rev.\ Lett.\  {\bf 113} (2014) 151601,
  arXiv:1406.6482 [hep-ex].
  %%CITATION = ARXIV:1406.6482;%%  
  
\bibitem{Abdesselam:2016llu}
  A.~Abdesselam {\it et al.} [Belle Collaboration],
 %{``Angular analysis of $B^0 \to K^\ast(892)^0 \ell^+ \ell^-$,''}
  arXiv:1604.04042 [hep-ex].
  %%CITATION = ARXIV:1604.04042;%%
    
\bibitem{Wehle:2016yoi}
  S.~Wehle {\it et al.} [Belle Collaboration],
%{``Lepton-Flavor-Dependent Angular Analysis of $B\to K^\ast \ell^+\ell^-$,''}
  Phys.\ Rev.\ Lett.\  {\bf 118} (2017) no.11,  111801
  % doi:10.1103/PhysRevLett.118.111801
  [arXiv:1612.05014 [hep-ex]].
  %%CITATION doi:10.1103/PhysRevLett.118.111801;%%    

\bibitem{models}
A partial list can be found in \cite{Capdevila:2017bsm} for example.

%%%%%%%%%%%%


%\cite{Capdevila:2017bsm}
\bibitem{Capdevila:2017bsm} 
  See for example B.~Capdevila, A.~Crivellin, S.~Descotes-Genon, J.~Matias and J.~Virto,
  %``Patterns of New Physics in $b\to s\ell^+\ell^-$ transitions in the light of recent data,''
  arXiv:1704.05340 [hep-ph], and references therein.
  %%CITATION = ARXIV:1704.05340;%%
  %28 citations counted in INSPIRE as of 21 Jun 2017

%%%%%%seesaw model citations go here

\bibitem{seesaw}
%\bibitem{Minkowski:1977sc}
  P.~Minkowski,
  %``Mu $\to$ E Gamma At A Rate Of One Out Of 1-Billion Muon Decays?,''
  Phys.\ Lett.\  B {\bf 67}, 421 (1977);
  %%CITATION = PHLTA,B67,421;%%
T.~Yanagida, in {\it Proceedings of the Workshop on the Unified Theory and the Baryon Number in
the Universe}, edited by O.~Sawada and A.~Sugamoto (KEK, Tsukuba, 1979), p.~95;
%  T.~Yanagida,
  %``Horizontal gauge symmetry and masses of neutrinos,''
  Prog.\ Theor.\ Phys.\  {\bf 64}, 1103 (1980);
  %%CITATION = PTPKA,64,1103;%%
M.~Gell-Mann, P.~Ramond, and R.~Slansky,
%``Complex Spinors And Unified Theories,''
in {\it Supergravity}, edited by P.~van Nieuwenhuizen and D.~Freedman
(North-Holland, Amsterdam, 1979), p.~315;
%\bibitem{Ramond:1979py}
  P.~Ramond,
  %``The Family Group in Grand Unified Theories,''
  arXiv:hep-ph/9809459;
  %%CITATION = HEP-PH/9809459;%%
S.L.~Glashow, in {\it Proceedings of the 1979 Cargese Summer Institute on Quarks and Leptons},
edited by M.~Levy {\it et al}. (Plenum Press, New York, 1980), p.~687;
%\bibitem{Mohapatra:1979ia}
  R.N.~Mohapatra and G.~Senjanovic,
  %``Neutrino mass and spontaneous parity nonconservation,''
  Phys.\ Rev.\ Lett.\  {\bf 44}, 912 (1980);
  %%CITATION = PRLTA,44,912;%%
%\bibitem{seesaw12}
J.~Schechter and  J.W.F.~Valle,
  % Neutrino Masses in SU(2) x U(1) Theories  %
  Phys.\  Rev.\ D {\bf 22}, 2227 (1980);
%%CITATION = PHRVA,D22,2227;%%
%J.~Schechter and  J.W.F.~Valle,
  % Neutrino Decay and Spontaneous Violation of Lepton Number %
  Phys.\  Rev.\ D {\bf 25}, 774 (1982).
%%CITATION = PHRVA,D25,774;%%


%%%%%%%%%%%

\bibitem{Langacker:1988vz}
% \cite{Langacker:1988vz}
P.~Langacker, S.~Uma Sankar and K.~Schilcher,
%``K-L $\to$ Mu E In SU(2)-L X U(1) And SU(2)-L X SU(2)-R X U(1) Models With
%Large Neutrino Masses,''
Phys.\ Rev.\ D {\bf 38}, 2841 (1988); 


\bibitem{Gagyi-Palffy:1994hx}
% \cite{Gagyi-Palffy:1994hx}
Z.~Gagyi-Palffy, A.~Pilaftsis and K.~Schilcher,
%``Heavy neutrino chirality enhancement of the decay K(L) $\to$ e mu in
%left-right symmetric models,''
Phys.\ Lett.\ B {\bf 343}, 275 (1995)
[arXiv:hep-ph/9410201]; 

  
  %\cite{He:2004wr}
\bibitem{He:2004wr} 
  X.~G.~He, G.~Valencia and Y.~Wang,
  %``Lepton flavor violating tau and B decays and heavy neutrinos,''
  Phys.\ Rev.\ D {\bf 70}, 113011 (2004)
  doi:10.1103/PhysRevD.70.113011
  [hep-ph/0409346].
  %%CITATION = doi:10.1103/PhysRevD.70.113011;%%
  %12 citations counted in INSPIRE as of 21 Jun 2017
  
    %\cite{Fujihara:2005uq}
\bibitem{Fujihara:2005uq} 
  T.~Fujihara, S.~K.~Kang, C.~S.~Kim, D.~Kimura and T.~Morozumi,
  %``Low scale seesaw model and lepton flavor violating rare B decays,''
  Phys.\ Rev.\ D {\bf 73}, 074011 (2006)
  doi:10.1103/PhysRevD.73.074011
  [hep-ph/0512010].
  %%CITATION = doi:10.1103/PhysRevD.73.074011;%%
  %10 citations counted in INSPIRE as of 21 Jun 2017

  
    % \cite{Inami:1980fz}
\bibitem{Inami:1980fz}
T.~Inami and C.~S.~Lim,
%``Effects Of Superheavy Quarks And Leptons In Low-Energy Weak Processes K(L)
%$\to$ Mu Anti-Mu, K+ $\to$ Pi+ Neutrino Anti-Neutrino And K0 <---> Anti-K0,''
Prog.\ Theor.\ Phys.\  {\bf 65}, 297 (1981)
[Erratum-ibid.\  {\bf 65}, 1772 (1981)].
%%CITATION = PTPKA,65,297;%%

% \cite{Buchalla:1995vs}
\bibitem{Buchalla:1995vs}
G.~Buchalla, A.~J.~Buras and M.~E.~Lautenbacher,
%``Weak Decays Beyond Leading Logarithms,''
Rev.\ Mod.\ Phys.\  {\bf 68}, 1125 (1996)
[arXiv:hep-ph/9512380].
%%CITATION = HEP-PH 9512380;%%


%\cite{Olive:2016xmw}
\bibitem{Olive:2016xmw} 
  C.~Patrignani {\it et al.} [Particle Data Group],
  %``Review of Particle Physics,''
  Chin.\ Phys.\ C {\bf 40}, no. 10, 100001 (2016).
  doi:10.1088/1674-1137/40/10/100001
  %%CITATION = doi:10.1088/1674-1137/40/10/100001;%%
  %1023 citations counted in INSPIRE as of 21 Jun 2017
  
  %\cite{Antusch:2006vwa}
\bibitem{Antusch:2006vwa} 
  S.~Antusch, C.~Biggio, E.~Fernandez-Martinez, M.~B.~Gavela and J.~Lopez-Pavon,
  %``Unitarity of the Leptonic Mixing Matrix,''
  JHEP {\bf 0610}, 084 (2006)
  doi:10.1088/1126-6708/2006/10/084
  [hep-ph/0607020].
  %%CITATION = doi:10.1088/1126-6708/2006/10/084;%%
  %273 citations counted in INSPIRE as of 23 Jun 2017
  
  %\cite{Fernandez-Martinez:2016lgt}
\bibitem{Fernandez-Martinez:2016lgt} 
  E.~Fernandez-Martinez, J.~Hernandez-Garcia and J.~Lopez-Pavon,
  %``Global constraints on heavy neutrino mixing,''
  JHEP {\bf 1608}, 033 (2016)
  doi:10.1007/JHEP08(2016)033
  [arXiv:1605.08774 [hep-ph]].
  %%CITATION = doi:10.1007/JHEP08(2016)033;%%
  %30 citations counted in INSPIRE as of 28 Jun 2017
  
  %%%%%%%%light nus for R(D) 
  
  %\cite{Bozek:2010xy}
\bibitem{Bozek:2010xy} 
  A.~Bozek {\it et al.}  [Belle Collaboration],
  %``Observation of B+ -> Dbar*0 tau+ nu_tau and Evidence for B+ -> Dbar^0 tau+ nu_tau at Belle,''
  Phys.\ Rev.\ D {\bf 82}, 072005 (2010)
  [arXiv:1005.2302 [hep-ex]].
  %%CITATION = ARXIV:1005.2302;%%  
  
 
%\cite{Lees:2012xj}
\bibitem{Lees:2012xj}
  J.~P.~Lees {\it et al.} [BaBar Collaboration],
  %``Evidence for an excess of $\bar{B} \to D^{(*)} \tau^-\bar{\nu}_\tau$ decays,''
  Phys.\ Rev.\ Lett.\  {\bf 109}, 101802 (2012)
  doi:10.1103/PhysRevLett.109.101802
  [arXiv:1205.5442 [hep-ex]].
  %%CITATION = doi:10.1103/PhysRevLett.109.101802;%%

%\cite{Aaij:2015yra}
\bibitem{Aaij:2015yra}
  {LHCb} Collaboration,
    %``Measurement of the ratio of branching fractions $\mathcal{B}(\bar{B}^0 \to D^{*+}\tau^{-}\bar{\nu}_{\tau})/\mathcal{B}(\bar{B}^0 \to D^{*+}\mu^{-}\bar{\nu}_{\mu})$,''
  Phys.\ Rev.\ Lett.\  {\bf 115}, no. 11, 111803 (2015)
  Addendum: [Phys.\ Rev.\ Lett.\  {\bf 115}, no. 15, 159901 (2015)]
  doi:10.1103/PhysRevLett.115.159901, 10.1103/PhysRevLett.115.111803
  [arXiv:1506.08614 [hep-ex]].
  %%CITATION = doi:10.1103/PhysRevLett.115.159901, 10.1103/PhysRevLett.115.111803;%%


%\cite{He:2012zp}
\bibitem{He:2012zp} 
  X.~G.~He and G.~Valencia,
  %``$B$ decays with $\tau$ leptons in nonuniversal left-right models,''
  Phys.\ Rev.\ D {\bf 87}, no. 1, 014014 (2013)
  doi:10.1103/PhysRevD.87.014014
  [arXiv:1211.0348 [hep-ph]].
  %%CITATION = doi:10.1103/PhysRevD.87.014014;%%
  %35 citations counted in INSPIRE as of 21 Jun 2017
  
    %\cite{Cvetic:2017gkt}
\bibitem{Cvetic:2017gkt} 
  G.~Cvetic, F.~Halzen, C.~S.~Kim and S.~Oh,
  %``Anomalies in (Semi)-Leptonic $B$ Decays $B^{\pm} \to \tau^{\pm} \nu$, $B^{\pm} \to D \tau^{\pm} \nu$ and $B^{\pm} \to D^* \tau^{\pm} \nu$, and Possible Resolution with Sterile Neutrino,''
  arXiv:1702.04335 [hep-ph].
  %%CITATION = ARXIV:1702.04335;%%
  %5 citations counted in INSPIRE as of 21 Jun 2017

%%%%%%%%%%%%%%%%

  
\end{thebibliography}
\end{document}